# The Rise and Possible Decline of Societal Complexity

Theodore Modis

**Abstract:** Societal complexity may be at a historical peak. Distinct from entropy, complexity tends to rise as systems move away from order, crest at an intermediate state, and decline as entropy continues increasing. The use of a thermodynamic analogy and the timing of major technological milestones—from fire to artificial intelligence—shows that the acceleration and recent compression of transformative events fit the derivative of a logistic growth curve. This pattern suggests that the rapid rise in structural and technological novelty may soon begin slowing. Notably, the trajectory parallels the bell-shaped rate of global population growth, consistent with the view that demographic expansion fuels innovation. If complexity growth is indeed cresting, societies face the challenge of managing heightened fragility while adapting to diminishing returns in transformative change. This perspective explores whether the rapid acceleration of technological innovation observed in recent centuries may reflect a civilizational system approaching the region of maximal complexity often associated with the edge of chaos.

---

Strategic Forecasting, Growth Dynamics, Lugano, Switzerland.
Email: tmodis@growth-dynamics.com

## Introduction

In the rigorous analysis of evolutionary systems, a clear distinction must be maintained between entropy and complexity—variables that are frequently conflated but occupy divergent trajectories. Entropy, fundamentally a measure of the number of microstate configurations consistent with a macrostate, follows the second law of thermodynamics. In closed and non-equilibrium systems alike, entropy typically exhibits a monotonic, S-shaped trajectory—characterized typically by a logistic growth function—toward an asymptotic maximum.

Complexity, conversely, represents the information-theoretic density or the difficulty involved in describing a system's configuration. It is not monotonically increasing with entropy or "disorder" (*1*). Unlike entropy, complexity follows a bell-shaped distribution. It thrives exclusively in the "sweet spot" or phase transition between total stochasticity (high entropy) and rigid order (low entropy). As a system approaches thermodynamic equilibrium and becomes uniform, its complexity approaches zero, regardless of its entropy level.

This principle has been formalized in different guises. In his book *The Quark and the Jaguar: Adventures in the Simple and the Complex*, Murray Gell-Mann argues that there is a trade-off between entropy and complexity, and that as entropy increases, complexity may increase only up to a certain point beyond which the system becomes too disordered to sustain its complexity (*2*). In their book *Into the Cool: Energy Flow, Thermodynamics, and Life*, Eric Schneider and Dorion Sagan propose the idea of increasing and decreasing complexity in relation to entropy (*3*). They argue that in complex systems as entropy increases, there is an initial increase in complexity, but eventually the system becomes too disordered and its complexity breaks down. In dynamical systems both periodic and random processes are considered simple, while complex and chaotic processes lie in between (*4*). Huberman and Hogg demonstrate that in discrete systems complexity takes low values for both ordered and disordered states while increases for intermediate states, tracing out an almost regular bell shape (*5*). In the same direction Kauffman has coined the term “complexity catastrophe” to explain the low complexity of an overly connected network similar to that of a sparsely connected network (*6,7*). In his book *The Big Picture: On the Origins of Life, Meaning, and the Universe Itself* Sean Carroll argues that complexity is related to entropy and that "complexity is about to begin declining" (*8*).Expecting decreasing complexity sometime in the future is a conclusion corroborated by Magee and Devezas who studied shorter-timescale technologically-driven or simply human-driven profound societal changes (*9*).

Could the rapid technological acceleration of the past few centuries represent a civilizational system operating near the edge of chaos? Many complex systems exhibit their richest dynamics near a boundary between order and randomness, often referred to as the "edge of chaos," a regime where systems exhibit maximal complexity and adaptability. Studies of biological evolution, neural networks, and artificial life suggest that systems operating in this regime can generate diverse structures while remaining sufficiently organized to sustain them. Work by Stuart Kauffman (*6*), Christopher Langton (*10*), and others has explored how adaptive systems may evolve toward such intermediate states. The historical acceleration of technological innovation in human societies may reflect a comparable phenomenon at a civilizational scale, in which expanding networks of interaction increase the diversity of possible structures while maintaining sufficient coherence for those structures to persist.

The aim of this article is not to provide a definitive measure of societal complexity, which remains difficult to quantify, but rather to explore whether several independent indicators— including simple physical analogies and historical patterns of technological change—are consistent with the possibility that complexity follows a bell-shaped trajectory during the long-term evolution of human systems.

## Conceptual Models of Complexity Dynamics

The divergence between Entropy and Complexity is best illustrated through the "Coffee and Cream Automaton" and the "Fair Die" stochastic model.

### *The Coffee Automaton*

Sean Carroll offers the metaphor of a cup of coffee with cream (*11*). In the initial state, where the cream sits undisturbed on top of the coffee, the system is ordered: entropy is low, and complexity is low because the state is simple to describe. In the final state of maximum entropy, the mixture is uniform and again easy to characterize. It is during the middle of the mixing process, when swirling, unpredictable tendrils of cream and coffee intertwine, that complexity reaches its peak. Carroll and two collaborators attempted to explore this idea quantitatively using a stylized version of the coffee-and-cream scenario, "Quantifying the rise and fall of complexity in closed systems: The coffee automaton" (*12*); however that study remains unpublished to date.

### *The Fair Six-Sided Die*

A simpler example—for which entropy and complexity can be easily calculated—demonstrates the same principle: a fair six-sided die in the very long run (*13*). As the die is thrown repeatedly over centuries, wear and tear gradually round its corners. Initially, the system's complexity increases because the rounding creates a seventh possible outcome—landing on a rounded apex—making the system harder to predict. However, as the die degrades into a perfect sphere, it lands the same way every time. Once again, the system becomes simple to describe: maximum entropy, minimal complexity.

## Complexity as the Derivative of Entropy

In the die example, entropy ($S$) is defined as "the information content," and complexity ($C$) as "the capacity to incorporate information." With these definitions it follows that entropy results from the accumulation (the integral) of complexity over time (*14*).

A profound mathematical relationship can thus be deduced between these quantities when analyzed through the lens of non-equilibrium thermodynamics. Complexity is treated as the rate of change—the derivative—of entropy:

$$C \cong \frac{dS}{dt}$$

This relationship stems from the particular information-related definitions used and cannot be rigorously generalized in all cases. In a state of thermodynamic equilibrium, for instance, entropy no longer changes with time, which would imply zero complexity. But all definitions of entropy and complexity share conceptual and/or mathematical connections (*15*). Therefore this relationship offers general qualitative insight. Indeed, systems in equilibrium exhibit very low complexity.

Grandy has also argued similarly from another angle: "... in nonequilibrium thermodynamics it is the time derivative of entropy that governs the ongoing macroscopic processes" (*16*).

Complexity reaches its zenith at the entropy inflection point. This point—where entropy's growth rate crests—marks the onset of information's degradation from structured meaning into noise. At this juncture, the system is undergoing the most intense structural reconfiguration and possesses the highest degree of "interestingness." Interestingness peaks when complexity is highest.

As the system's entropy keeps growing asymptotically toward it final maximum the information content may technically remain high in an archival sense, but without the dynamic rate of change provided by the derivative, it degrades into uninteresting, random data as the system settles toward equilibrium.

### Quantifying Complexity via Historical Milestones

While it's straightforward to calculate the complexity of a fair die, quantifying the complexity of more complex systems, like the coffee-and-cream example, becomes daunting. The challenge is even worse with human beings and their environment. How do we measure how complex a human is? Is a vegetable more complex than a computer? No universal definition or direct measurement method for complexity exists.

One possible indicator of societal complexity is the rate at which major technological or cultural innovations appear. Each such milestone introduces new structures, interactions, and capabilities within human systems. While no single metric fully captures societal complexity, the temporal spacing of major innovations may provide a useful proxy for the rate at which complexity is generated.

There has been a number of quantitative studies of the accelerating rate of change—technological or otherwise (*17-19*). A recent analysis examines the 14 most important milestones tied to human development, beginning with the domestication of fire and ending with AI (*20*). A sequence of widely recognized technological milestones spanning the development of human civilization has thus been distilled, see Table I. These events were selected because they represent widely acknowledged transformations that substantially expanded the scale or diversity of human interactions. To offset the inherent subjectivity in choosing milestones, inputs have been corroborated by world-renowned scientists, such as Nobel laureates in the sciences, and other reputable sources.

This milestone-based approach is necessarily approximate and intended primarily as a heuristic indicator rather than a precise quantitative measure of societal complexity.

Rather than focusing on the dates themselves, the analysis considers the intervals between successive milestones, which compresses dramatically over time. Hundreds of millennia separate fire and language; only decades separate the Internet and AI, with a slowing down of the time compression noticeable recently.

This pattern resembles the dynamics of systems approaching a peak in complexity. In many complex systems, the rate at which new structures form increases as the system moves away from highly ordered states and enters regions where interactions between components become more numerous. However, once systems approach high levels of disorder, the capacity for maintaining organized structure may decline. Under such conditions, complexity can follow a bell-shaped trajectory: increasing during an intermediate phase before eventually decreasing as entropy continues to rise.

If societal complexity follows a similar trajectory, the observed acceleration of technological milestones may represent the ascending portion of such a curve. In this interpretation, the rapid pace of innovation during recent centuries reflects the expansion of interconnected human networks and the increasing diversity of technological components. Whether this acceleration will continue indefinitely, plateau, or eventually decline remains an open question, but the historical pattern suggests that the dynamics of societal complexity may be comparable to those observed in other complex systems.

**Table I - The 14 milestones with the time before the year 2000 noted in years**

| No. | Milestone | Years Before 2000 |
|---|---|---|
| 1 | Domestication of fire | 700,000 |
| 2 | Emergence of *Homo sapiens*, acquisition of spoken language | 300,000 |
| 3 | Earliest burial of the dead | 100,000 |
| 4 | Earliest rock art | 40,000 |
| 5 | First cities, invention of agriculture | 10,070 |
| 6 | Development of the wheel, writing | 5,450 |
| 7 | Classical thought: Buddha, democracy, city states | 2,536 |
| 8 | Numerical systems: zero, decimals | 1,415 |
| 9 | Renaissance, printing press, scientific method | 487 |
| 10 | Industrial revolution (steam engine, Enlightenment era) | 200 |
| 11 | Modern physics (widespread development of science and technology: electricity, telephone, radio, automobile, plane, etc.) | 97 |
| 12 | Nuclear energy, DNA, transistor | 51 |
| 13 | Internet, genome sequencing | 5 |
| 14 | Artificial Intelligence, big data, social media | -23 |

Each milestone is of maximum importance, which depends on two factors: the magnitude of complexity it introduces and the duration of its influence before the next milestone occurs. The greater the complexity jump at a given milestone, or the longer the ensuing stasis, the greater the milestone's importance will be. This gives rise to a simple formula:

$$Importance \propto Complexity \times Duration$$

In the approximation that milestones of *maximum* importance can be considered to be of *equal* importance ($I$), the increase in complexity ($\Delta C_i$) associated with milestone $i$ can be estimated inversely from the time gap ($\Delta T_i$), the time between milestone $i$ and milestone $i$+1:

$$\Delta C_i = \frac{I}{\Delta T_\mathrm{i}}$$

Complexity values can thus be calculated in arbitrary units with error bars reflecting the spread of constituent events within a milestone. The curve is steeply rising. The steep ascent began around the Renaissance, quickened through industrialization, and possibly crested in the digital age around the early twenty-first century—our times. The equal-importance approximation is necessarily a simplification, but it is defensible: the milestones were selected precisely because each was transformative enough to reshape subsequent human development, making them broadly comparable in civilizational weight even if not identical in scale.

Entropy’s S-shaped trajectory with an inflection point around the middle resembles a natural-growth curve often described by the *logistic* function:

$$f(x) = \frac{M}{(1 + e^{-\alpha(x - x_0)})}$$

where M, α, and $x_0$ constants, and $x$ the time variable.

Its derivative resembles a bell-shaped curve, often referred to as a life cycle, and peaks at the inflection point, when entropy grows fastest:

$$f'(x) = \frac{M\alpha}{(1 + e^{-\alpha(x - x_0)})(1 + e^{\alpha(x - x_0)})}$$

Given that entropy follows an S-shaped pattern and that complexity tracks the derivative of entropy, it is reasonable to fit the estimated complexity values of the milestones with this bell-shaped derivative of a logistic function. The fit yields a good match to the data when $x$ represents the sequential milestone number. The fitted curve aligns closely with all data points, see Figure 1.

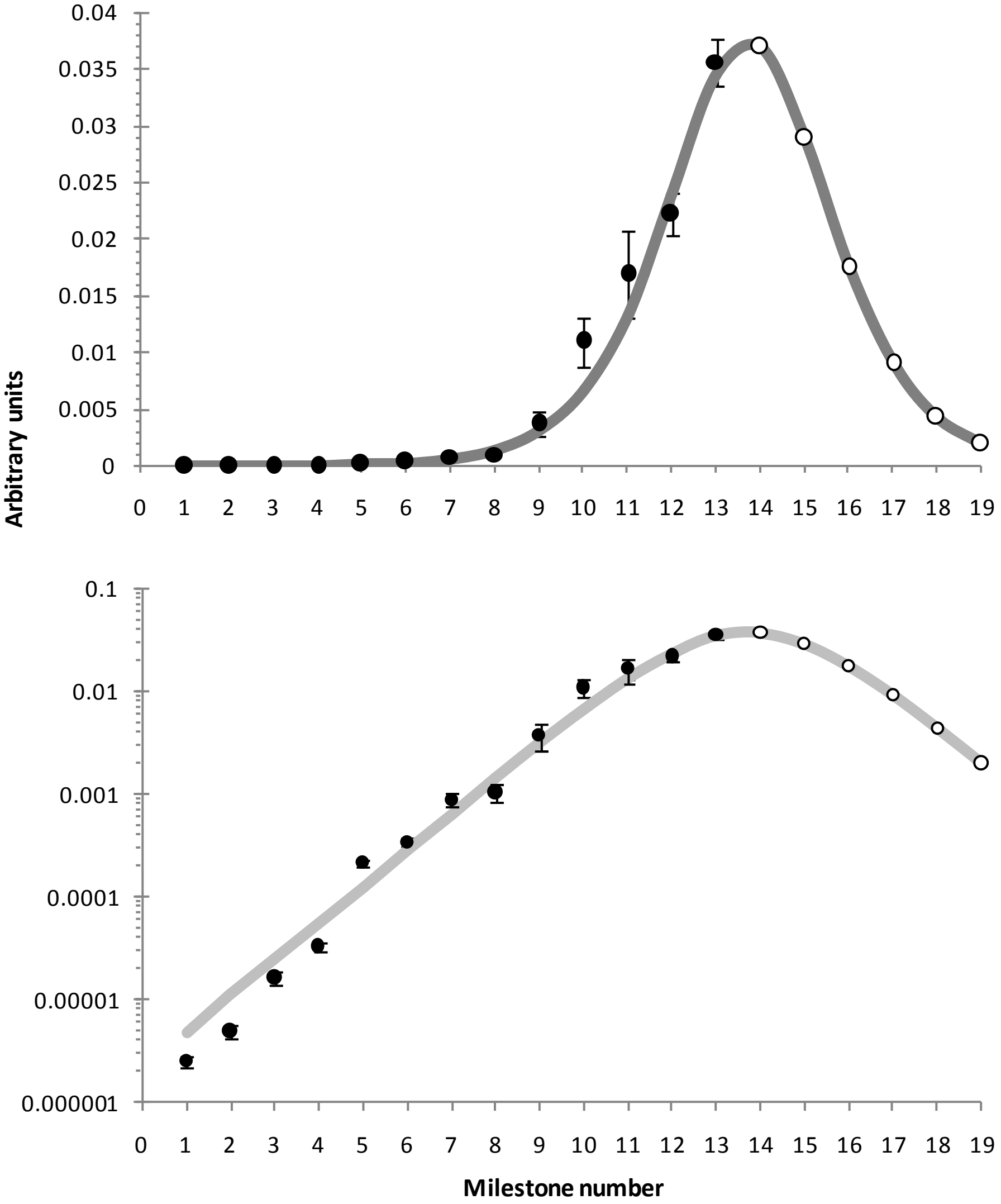

**Figure 1. Complexity.** The thick gray line is a logistic life-cycle fit to the data of the first thirteen milestones. The vertical axis depicts the change in complexity, with logarithmic scale in the lower graph. The little white circles on the extrapolated trend indicate the expected complexity of future milestones; the first one—Number 14—refers to AI.

The goodness of the fit has been evaluated with a Graphical Analysis of Residuals, plotting the fit vs. the data with a trend line:

$$Y = 1.0174 * X - 0.00049 \text{ with } R^2 = 0.980$$

which indicates good accuracy ($R^2 \approx 1$), no systematic bias (intercept ≈ 0), and no data-dependent bias (slope ≈ 1), see Table II.

**Table II - Fit Results**

| Function fitted | | | | Goodness of Fit | | |
|---|---|---|---|---|---|---|
| | $\alpha$ | $M$ | $x_0$ | $R^2$ | Slope | Intercept |
| $\frac{M\alpha}{(1+e^{-\alpha(x-x_0)})(1+e^{\alpha(x-x_0)})}$ | 0.8119 | 0.1849 | 13.71 | 0.980 | 1.0174 | 0.00049 |

AI is positioned at the end of complexity's maximum. Its impact is forecasted to exceed that of the Internet, representing the final "steep" climb of the curve. The model identifies a subsequent milestone cluster around the year 2050—likely comprising bioengineering, neuroscience, nanotechnology, and quantum computing. This cluster is expected to add less complexity than AI, but surpass the combined complexity impact of the 20$^{th}$-century triad (Milestone #12: Nuclear Energy, Transistor, and DNA). It represents the onset of diminishing returns. Subsequent innovations will yield decreasing "true new structure" as the derivative of the curve begins its inevitable descent.

### Population Growth as a Complexity Driver

Population dynamics seem related to the bell shape of the complexity trajectory in more than one way. With an average life expectancy of 80 years, the baby-boom generation—defined as 1934 to 1972 (*20*)—straddles almost perfectly the peak of the complexity curve. Whether by coincidence or consequence baby boomers have lived through the most complex period in the evolution of human society. These will likely be the most interesting times in human evolution.

But there appears to be a more significant connection between world-population dynamics and the bell pattern of complexity. Since 1950, the growth of world population has been meticulously tracked, and the data align remarkably well with a logistic growth curve, projecting around 9 billion by 2040, approaching stabilization (*20*). What's particularly intriguing is the behavior of the rate of change of the population growth. It reaches a maximum in 1997, forming a bell-shaped pattern remarkably similar to the complexity curve. When plotted together, the population growth rate leads the complexity curve—plotted as a function of time—by approximately 23 to 25 years, see Figure 2. This temporal offset mirrors the approximate time it takes for a newborn to reach maturity and begin making significant contributions to societal complexity.

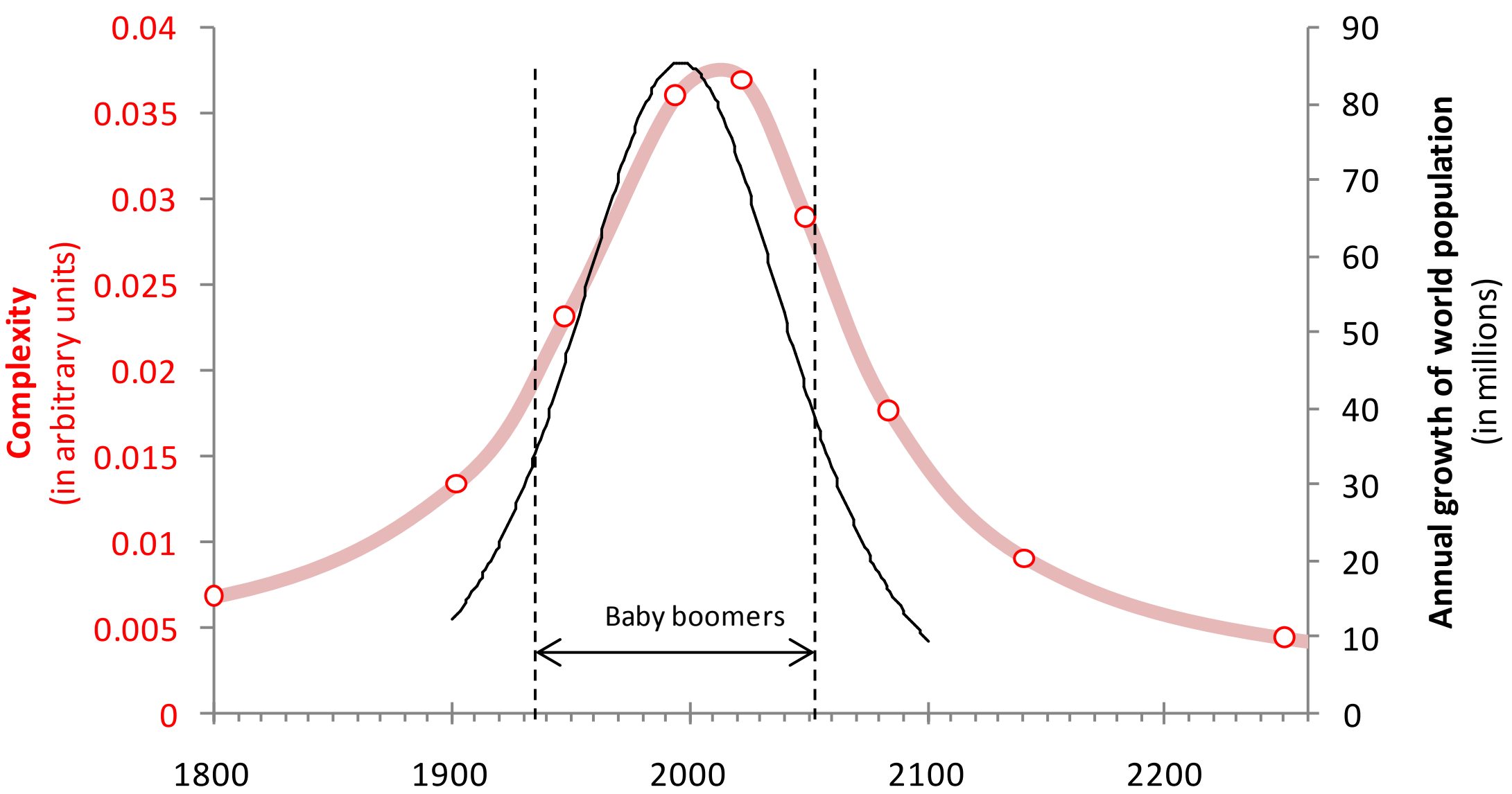

**Fig. 2. Complexity and population growth.** The rate of the world-population growth (black line) precedes the complexity curve (red line) by 23-25 years. The x-axis shows calendar years; the left y-axis shows complexity in arbitrary units; the right y-axis shows annual population increments in millions. The open circles designate the timing of milestones, with AI at the highest point.

This alignment doesn't amount to proof of causation, but it supports a compelling qualitative argument. As the population expands, so does the overall entropy of the human system. And since complexity tends to track entropy's rate of change, it is plausible that population dynamics could influence how complexity evolves.

Economist Robin Hanson has argued that population decline implies innovation decline (*21*). If we take innovation as a proxy for complexity, his argument may be rephrased to: A decline in population growth rate implies a decline in complexity 25 years later. Population drives interactions; interactions drive complexity; both waves crest together.

Complexity in social and technological systems has been widely studied within the framework of complex systems theory. Work by Yaneer Bar-Yam (*22*) and others has emphasized that complexity can be understood in terms of the information required to describe interactions among system components across multiple scales. As the number and diversity of such interactions increase, systems may exhibit increasingly rich structural organization. The historical acceleration of technological innovation may reflect the expansion of such interactions within human networks.

But an eventual decrease has been suggested as early as 1972 with the report "Limits to Growth" by the club of Rome, which predicted decreases in both population and industrial capacity (*23*).

### Managing the Peak

Living at the top of the bell may be the most stimulating yet precarious position human society can occupy. Novelty abounds, but so does instability. Every new connection creates potential cascades of failure; every new capability demands ethical restraint. Managing complexity becomes civilization's central task.

One implication is clear: biological and societal systems have flourished during increasing complexity. They may fare less well during declining complexity. To delay this decline and prolong conditions under which humans and civilizations prosper society may benefit from efforts to flatten the complexity curve.

Natural growth processes are notoriously resistant to manipulation. Still, changes can occur if introduced gradually, adiabatically. Slowing down may be the way to reduce complexity's decline. Embracing simpler lifestyles, intentional living, and reduced consumption could temper the accelerating churn of technological and societal transformations.

Signs of self-correction are already visible. Movements toward minimalism, slow living, and "degrowth" challenge the cult of acceleration, reflecting a growing intuition that not all progress is defined by speed. Though still marginal, these movements may represent the early stirrings of a cultural shift—an unconscious attempt by society to flatten complexity's curve in order to extend the longevity of an environment in which humanity traditionally thrived.

Complexity, like entropy, has a direction—but unlike entropy, it can fall. Recognizing where humanity stands on this curve is the first step toward navigating what comes next.

**Funding:** No funding.

**Competing interests:** No competing interests.

### AI Statement

I used AI tools for light editorial assistance (language refinement and structural feedback), but the argument, research, and authorship are entirely my own.